\documentclass[sigconf, anonymous=false, review=false ]{acmart}
\renewcommand\footnotetextcopyrightpermission[1]{} 
\usepackage{booktabs} 
\usepackage{grumble}
\setcopyright{none}

\acmDOI{10.475/123_4}
\acmISBN{123-4567-24-567/08/06}
\acmConference[MARS'17]{The 7th Workshop on Multi-core and Rack Scale Systems}
{April 2017}{Belgrade, Serbia}
\acmYear{2017}
\copyrightyear{2017}
\acmPrice{15.00}

\begin{document}
\title{Making data center computations fast, but not so furious}


\author{Daniel Porto}
\affiliation{%
  \institution{INESC-ID/IST -- U.\ Lisboa}
}

\author{Jo\~{a}o Loff}
\affiliation{%
  \institution{INESC-ID/IST -- U.\ Lisboa}
}

\author{Rui Duarte}
\affiliation{%
  \institution{INESC-ID/IST -- U.\ Lisboa}
}

\author{Luis Ceze}
\affiliation{%
  \institution{University of Washington}
}

\author{Rodrigo Rodrigues}
\affiliation{%
  \institution{INESC-ID/IST -- U.\ Lisboa}
}

%

\renewcommand{\shortauthors}{D. Porto et al.}

\begin{abstract}
\daniel{make it two paragraphs}
\daniel{1st talks about the context (comp sprinting), 2nd opportunities (fault tolerance+overcloking )}


We propose an aggressive computational sprinting variant for data center environments. 
While most of previous work on computational sprinting focuses on maximizing
the sprinting process while ensuring non-faulty conditions, we take advantage of the existing
replication in data centers to push the system beyond its safety limits.
In this paper we outline this vision, we survey existing techniques for achieving it, and we present some design ideas for future work in this area.



\end{abstract}
\keywords{Computational Sprinting, Overclock, Fault tolerance, Data center.}

\maketitle
\disclaimer
\section{Introduction}
\daniel{Structure}
\daniel{1. Serviços de larga escala em datacenters usam replicação pra disponibilidade, }
\daniel{2. replicas, quando nao ocorre problema jogam o resulatado fora e gastam energia }
\daniel{3. Energia eh uma preocupação em data centers, facebook tinha  turbo boost (industry supported overclock) até ter um serviço pra fazer power cap}
\daniel{4. Sprinting também economiza energia, }
\daniel{5. Overcloking alem dos limites tem aplicações interessante em outras areas como FPGA mas tem como problemas a presença de faltas}
\daniel{6. Replicacao pode permitir elevar sprinting a um outro nível ao passo que mascara as faltas.}

\daniel{computational sprinting vs turbo boost:}
\daniel{Computational sprinting [27] utilizes thermal capacitance to temporarily overclock the CPU to achieve increased performance for a short period of time. This is different from TM in that TMcan achieve sustained overclocking if there are idle resources on the CPU.}
\daniel{ref Dynamic management of in modern multi-core chips TurboMode}

Today's global scale Internet services run in large data center infrastructures and are accessed by millions of users.  The sheer scale in which these systems
operate is such that the design of the infrastructure underlying data center
systems must expect an environment where faults are the norm, and
no longer the exception. 

A key technique for building systems to provide high availability
despite faults is to employ redundancy, often
through the use of distributed replication protocols. However, this
redundancy has a resource usage cost associated with it.
For instance,  Google uses 5 replicas for the F1 Advertising
Back-end~\cite{Corbett2012}, which multiplies the number of servers required to run this service by that factor. 

Furthermore, F1 is not an isolated example, since redundancy is present
in a large fraction of the systems that are part of the software stack
of major companies such as Google or Facebook~\cite{malte:thesis}.
Thus, redundantly storing data and performing computations in multiple
servers increases the energy cost and uses resources that otherwise
could be allocated to serve other types of requests.
Nevertheless, this expense is 
seen as an important insurance, because faults lead to service downtime which,
in turn, affects revenue~\cite{DeCandia2007}.



On an orthogonal direction, energy efficiency is also a concern for data center operators, as it impacts the requirements and consequently the cost of the infrastructure for power delivery. This is a pressing problem both because the power consumption of servers is increasing with the advances in density and number of cores, and because the cost
associated with an increase in power capacity can be very high, reaching tens of
thousands of USD per additional MegaWatt \cite{Fan2007}. 

Moreover, while over-subscription of
data centers' power supply allows for accommodating infrequent correlated spikes
in server power consumption, it exposes data centers to the risk of tripping
power breakers and causing outages. For instance, Facebook initially had to disable
dynamic overclocking (Intel {Turbo Boost}~\cite{IntelTurboBoost}) in one of its
clusters, due to an insufficient
power margin, despite the potential benefits in performance of this technology. 
To safely enable turbo mode while avoiding the risk of outages in high load periods, they designed a
power management system to cap energy consumption according to the data center
power budget ~\cite{Wu2016}.

To lower energy consumption while also maintaining and even improving performance, a recent approach called \textit{computational sprinting}~\cite{Raghavan2012}
exploits the \textit{thermal capacitance} of materials to activate cores
(parallel sprinting) or overclocking the CPU (frequency sprinting), thus exceeding the 
sustained cooling capabilities of the system for short periods. As a result, this scheme is
able to improve application responsiveness by up to 6x and save about 30\% on power
for a conventional Core i7 Desktop chip, as a consequence of finishing
computations in a shorter amount of time~\cite{Raghavan2013}. This technique was also extended to
data centers, in which more interactive workloads such as search or news feeds,
that exhibit occasional bursty behavior, can benefit from short
performance boosts~\cite{Zheng2015}.



Techniques like computational sprinting, that push the limits of
what the hardware is designed to do, are conservative with respect
to the safety of computations. This is mainly because exceeding hardware specifications
leads to system instability. For instance,
overclocking or activating a
large number of cores for long periods leads to overheating, and exceeding the
Thermal Design Power (TDP) of the circuit may cause
faults~\cite{Nightingale2011}, reduce the lifespan, or even physically damage the
chip~\cite{inteldatasheet}. Hence, after each sprint the CPU ought to switch to
a cool-off mode.

Notwithstanding, even outside stable configurations, overclocking has been
explored with interesting results. For instance, DSP-accelerators  implemented
with FPGAs can save up to 39\% on hardware resources by overclocking for the same
output quality~\cite{duarte2015}. It was also observed that as the frequency increases the errors in computations gradually appear in the output,
up to the point where the program stops producing meaningful results.

\daniel{shall wemention that intel new processors have better overclocking support???}
\daniel{shall we mention isolation techniques such as virtualization???}
\daniel{note also that a large fraction of the apple processors nowadays are
	composed by accelerators, ref: Computational sprinting}

In this paper, we envision bringing together these two vectors, by leveraging the redundancy 
that is already present in a large fraction of data center systems to safely push the limits of computational
sprinting in data center environments. In other words, our goal is to
aggressively explore overclocking settings,  
while taking advantage of the existing redundancy introduced by
fault tolerance protocols to systematically mask faults that surface,
in order to extend the benefits of sprinting (energy efficiency/performance).
Furthermore, we intend to explore the synergies and subtle
interactions between the two vectors.
For instance, 
the sprints can be coordinated in a way that a subset of the replicas uses
a more aggressive but unsafe sprinting to decrease the overall latency, but
there is also
a sufficient number of non-sprinted replicas that check the results and ensure both availability and correctness
in the case of faults. 

The remainder of the paper provides an overview of the key techniques that we can leverage as building blocks.


%



\section{Computational sprinting }
\daniel{
	parallel sprinting
	freqnecy spritnig (turbo boost)
	data center spritnig
	overclocking
}
\daniel{shall we mention the need of proper cooling?}
\daniel{shall we mention that activating cores increase power linearly while \
	and increasing frequency/voltage augment power quadractically? }

\daniel{intro - why sprinting?}
Advances in CMOS technology have enabled the design of modern multi-core
processors, packing an increasing density of transistors at each new generation.
However, more transistors implies an increase in power density, at
a rate which exceeds the ability to dissipate the heat generated~\cite{Raghavan2013}. As a
result, continuously operating all the processing units at full power can
permanently damage the chip due to overheating. Consequently, some of the
 cores must remain off most of the time, a limitation known as dark
silicon~\cite{Esmaeilzadeh2011}.

\daniel{how sprinting works}
While it is not possible to activate all the processor cores at once in a
sustainable way, there exist proposals for optimizing
performance within safe temperature and power limits.
These solutions (outlined next),
leverage the fact that after activating a sprint, the temperature of the
components does not rise instantaneously. Instead, it can take a few seconds
for the heat to propagate through the chip package, allowing certain workloads
to finish before it overheats. 

\daniel{describe parallel sprinting} 
\textit{Parallel sprinting}. 
This approach consists of activating various dark silicon cores for up to a
time limit (e.g., 1 sec.)
before deactivating cores to cool off. Parallel sprinting can be
optimized according to two policies: for maximum responsiveness, it activates
all cores at maximum frequency and voltage; for optimal energy efficiency, it activates all
cores at minimum frequency and voltage~\cite{Raghavan2012}. Parallel sprinting 
is particularly interesting for mobile devices since a large part of mobile processors 
are comprised of accelerators that are inactive most of the time. 
In addition, mobile applications normally have interactive workloads that
are characterized by short bursts and long idle times waiting for user input~\cite{Raghavan2013}.

Note that there is an interesting research question, which we intend to explore, 
of whether data center workloads also have such characteristics. 
However, even if they do not, we can still attempt to split replicas in an 
alternating fashion between a group of replicas that are sprinting and another group that are cooling off.

\daniel{describe frequency spritnig}
\textit{Frequency Sprinting}.
A concrete example of frequency sprinting is the dynamic overclocking
technology presented in commercial products such as Intel processors with
{Turbo Boost 2.0}~\cite{IntelTurboBoost}. In a nutshell, Turbo Boost increases the frequencies and voltages of processor cores above a
normal safe operation threshold for short periods.
The sprint frequency target is defined automatically according to the available resources,
allowing the processor to improve performance of both single and multi-threaded workloads.
The algorithm that controls when sprinting is activated, takes into account the current frequencies of processor cores, the temperature
of the package, and its power consumption~\cite{inteldatasheet}. 

\daniel{describe data center sprinting - multiserver}
\textit{Data center sprinting}.
Data centers can also experience bursty workloads, e.g., due to shared
resources, maintenance activities, garbage collection events, or spikes in service
popularity~\cite{Dean2013,Gmach2007}. Therefore, they might be a good match
for sprinting. Moreover, the dark silicon phenomenon is likely to be prevalent in data centers, 
as supported by predictions that stated that by 2024 more than 50\% of the
chip must be powered off~\cite{Esmaeilzadeh2012}.
A current approach called Dynamo~\cite{Wu2016}  employs power capping
to enable frequency sprinting while ensuring that the energy drawn remains within the
power budget limits. Alternatively,  \cite{Zheng2015} employs coordinated
parallel sprinting, using the existing data center backup power supply (e.g. batteries), 
 to provide the extra power for
both sprint and cooling.

\daniel{describe overcloking: less reliable/error prone  single server}
\textit{Overclocking}.
Pushing hardware to work beyond the prescribed frequency has been studied by
other research communities~\cite{Shi2014}. One approach is increasing the frequency, 
without sufficiently increasing the voltage, which can lead to faults, 
because it may violate the propagation delays of circuit
critical paths. An interesting characteristic of these fault patterns is 
that errors gradually appear in the output as
frequency increases.
This gradual slope in the fault behavior happens because circuit
designers are conservative in the estimates of path delays and keep a guard
margin between the estimated clock frequency and the reported maximum clock
frequency~\cite{duarte2012high}, opening the opportunity to explore these limits.

\section{Fault tolerance}
Data center systems often rely on distributed replication protocols
for operating correctly in the presence of faults. These protocols are
designed under certain assumptions about the environment, such as
fault behavior (e.g, crash, fail-stop, or Byzantine) and timing (e.g.,
the synchronous vs. the asynchronous model).  Making wrong
assumptions about the environment can either put the safety properties
of the system at stake or impose an unnecessary cost in terms of
replication and consequently energy and infrastructure.

A pragmatic choice for replicating
services are asynchronous crash fault tolerant (CFT)
protocols~\cite{part-time-parliament,Ongaro2014}, because they cover
the most common fault and timing behaviors. However, since aggressive
overclocking can lead to data corruption or errors in a computation,
CFT protocols may be too optimistic for sprinting. Byzantine fault
tolerant (BFT) protocols~\cite{Castro1999,Clement2009,Kapitza2012} are
able to capture data corruption, but they are pessimistic regarding the
behavior of faulty replicas.  In particular, a BFT adversary may
control a fraction of the replicas, allowing collusion, creating the
worst case attack scenario. 
Such pessimism leads to a higher replication costs.


Visigoth fault tolerance (VFT)~\cite{Porto2015} allows for calibrating the
fault tolerance of the system according to the deployment. While it can
be configured to capture only crashes with the same replication requirements of
CFT, VFT can also capture data corruption with a small additional cost.
When compared to BFT, the replication requirements of VFT grows
slower as the number of tolerated faults increases. This is because VFT
 assumes bounded collusion, i.e. the
number of replicas that deviate from their expected behavior in the same way simultaneously.
One of the reasons why this assumption may be realistic for overclocked systems is that the variations in the chip manufacturing process causes processors
to have different stability configurations~\cite{duarte2015}. Additionally, we can enforce such diversity by carefully controlling overclocking at different replicas.

Other models have been proposed along the same lines as VFT, namely
XFT~\cite{Liu2016}, which has the same replication requirements
as CFT while capturing Byzantine behavior, although not simultaneously
with asynchrony. As future work, we can also explore whether this
variant in the set of assumption is met in practice.



\daniel{Overclocking + fault tolerance}
\section{Conclusion}

\daniel{Data center sprinting ->  concept, data center scheduler (select
	hotspots, enable  sprinting for certain nodes, vantagens: straggler mitigation,
	leverage spare resources due to replication)}
\daniel{Distributd systems are designed to tolerate faults.
	fault tolerant protocols deployed in data centers mask faults
	byzantine fault tolerance, XFT, VFT}
\daniel{pratictioners/industry avoid overclocking due to instability (no unlocked xeon
	processors, but TurboBoost is already available) plus data center operators
	often have custom hardware (google)}

We presented our vision on the potential of combining computational sprinting and fault tolerance to enable higher savings on energy and improved performance
for replicated systems. We intend to explore challenges of this approach, by designing and implementing a system that explores the opportunities identified here.

\section*{Acknowledgments}

This research is funded by the European Research Council (ERC-2012-StG-307732) and by the FCT (UID/CEC/50021/2013).

\bibliographystyle{ACM-Reference-Format}
\bibliography{sigproc}

\end{document}